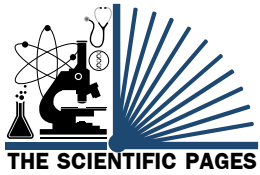

# Trends in Artificial Intelligence

# What Constitutes Elemental Shape Information for Biological Vision?

*Ernest Greene\* and Onyinye Onwuzulike*

*Laboratory for Neurometric Research, Department of Psychology, University of Southern California, Los Angeles, USA*

*The part itself would be seen not as a "part" of some earlier figure but as a self-sufficient whole in its own right* Friedrich Wulf [1]. We do not yet understand how the vertebrate visual system provides for recognition of objects. Countless experiments have been performed to examine the contribution of cues such as color, texture, and shadowing, but the most important cues are the contours of the outer boundary. Most objects that we can name can be identified as a silhouette, or equally well as a line drawing of the boundary. This has long been appreciated, so it is somewhat surprising that after more than a century of experimental research, we have not yet established how our visual system encodes this shape information.

Many might object to the last statement, for there is an abundance of research on how the visual system of higher vertebrate's registers lines and edges and countless discussions of how neurons could combine those responses for purposes of shape recognition. The prevailing views are so well known that no more than a brief reprise is needed before giving the reasons why they do not provide a satisfactory explanation.

Most theories pivot on the neurophysiological results reported by Hubel & Wiesel [2,3] and related assertions by Marr [4] in the field of machine vision. Both camps focused on the contours that are present in an object, and especially on the outer boundary. Hubel & Wiesel discovered that neurons in primary visual cortex (V1) selectively register the orientation of elongated bars. Further, these neurons showed optimal responding to oriented stimuli shortly after birth of the animal, so the development of orientation selectivity was not based on learning [5]. It was assumed that the elongated receptive field design of V1 neurons was based on anatomical convergence of optic tract fibers, mapping aligned receptive fields of retinal ganglion cells onto the orientation-selective neurons of V1. This hypothesis was supported by numerous laboratories, with especially definitive evidence being from [6].

Registering the contours of an object is now generally viewed as an elemental, i.e., essential, step toward shape recognition. As evidence accumulated that the neurons of inferotemporal cortex are involved in shape recognition, it was hypothesized that learning provides for effective convergence of connections from the V1 neurons to provide inferotemporal neurons with shape selectivity. In other words, recognition of a given shape would require training that modified the connectivity from V1 through the ventral pathway so that one or more neurons in inferotemporal cortex would be activated by a specific shape.

Numerous models, alternatively described as neural network or connectionist models, have been developed on the basis of this theory [7-14]. The models have various degrees of non-specific connectivity among the layers of model neurons as a starting point. Then training trials are applied to move or otherwise modify the strength of connections from one layer to the next to create selectivity of response by one or more neurons in the final layer. The goal is to teach the network to discriminate among various shapes, and to do so even if there are changes in location, size, or rotation of the shape to be identified.

Greene [15-17] has challenged these assumptions and models. For this work, shape boundaries were displayed as a string of dots using a 64 × 64 array of LEDs. The earlier experiments [15,16] examined recognition of known shapes, which included animals, vehicles, tools, furniture, and such. Recognition was still possible when the number of dots in the boundary was substantially reduced, even to the point where the span between adjacent dots was greater than the longest receptive fields of orientation-selective neurons in V1 [15]. Greene & Hautus [17] used an inventory of unknown shapes, and provided evidence of immediate above-chance recognition for sparse-dot versions of shapes that were seen only once. This indicates that our visual system does not require training to encode simple 2D shapes, which is at odds with the training requirements of connectionist models.

Wulf [1] presented observers with simple line drawings and later asked for whether the original stimulus could be remembered and reproduced when cued by a fractional version. He found that fractional versions would sometimes elicit shape perceptions that were unlike those elicited by the original stimulus. The findings reported here have much in common with his conclusions, and support the argument that orientation, length, and curvature of contours are not "elemental" shape cues.

The present work used the matching protocol used by Greene & Hautus [17], which is illustrated in Figure 1. This protocol displays unknown shapes that are designed to have little similarity to known objects. A given shape in the inventory consists of a string of dots that form a single continuous loop. Shapes that serve as "targets" are sampled at random from the inventory, and then may be reduced in dot-density by applying an algorithm that maximizes the spacing (as steps around the dot string) among the sampled dots.

Target shapes were displayed at either a 4% or 32%

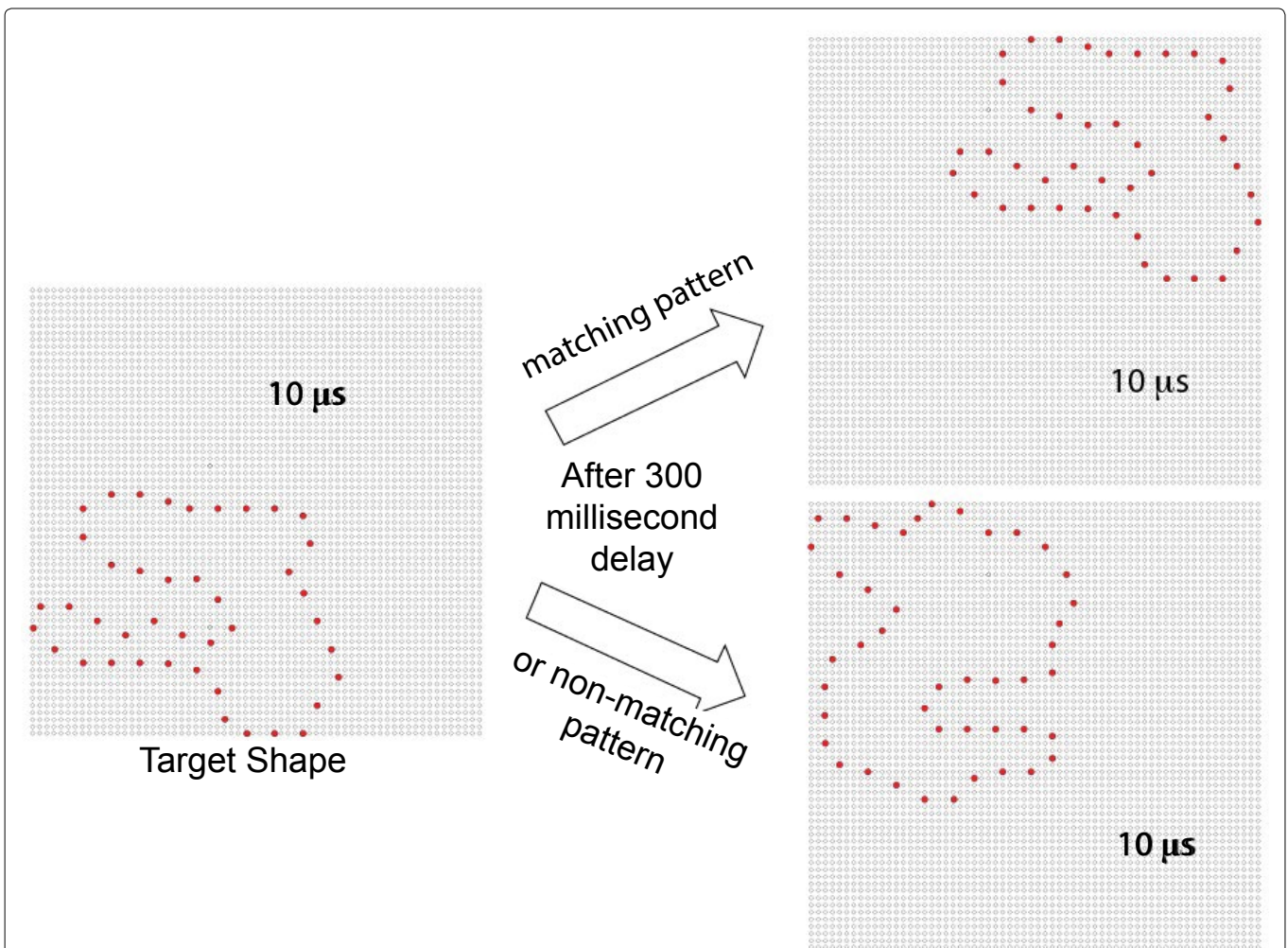

**Figure 1:** The matching task is illustrated. Target shapes were random selections from a 480-shape inventory, and were displayed at 4% or 32% density. A 32% density shape is illustrated here. Comparison shapes were displayed at densities that ranged from 4% to 32% in the opposite corner of the display board. The comparison shape was the same shape as the target

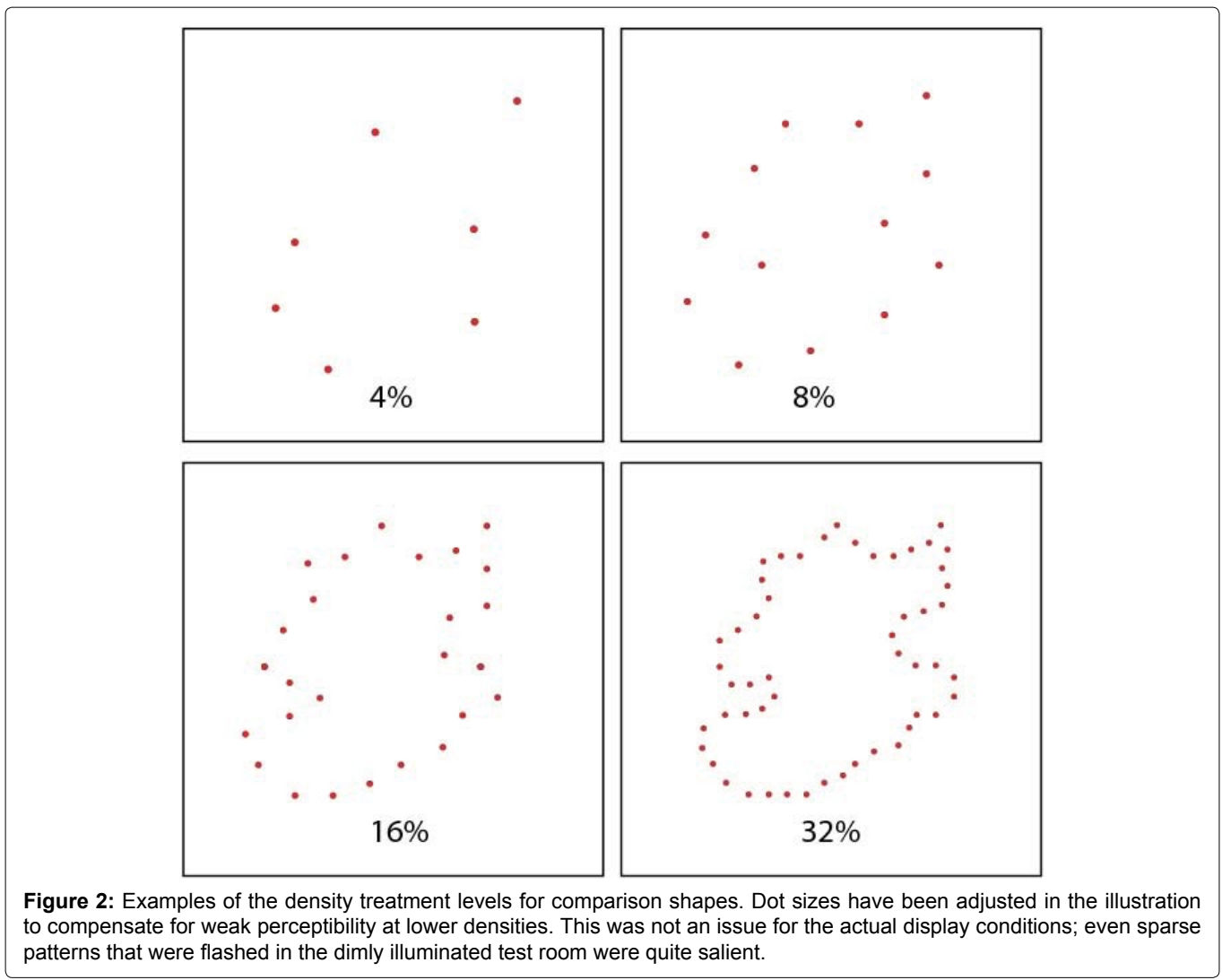

**Figure 2:** Examples of the density treatment levels for comparison shapes. Dot sizes have been adjusted in the illustration to compensate for weak perceptibility at lower densities. This was not an issue for the actual display conditions; even sparse patterns that were flashed in the dimly illuminated test room were quite salient.

density. Each target shape was displayed in one corner of the board with a 10-microsecond flash of all the boundary dots. After a 300-millisecond delay, a low-density comparison shape was flashed in the opposite corner. The comparison shape provided either a low-density subset of the target boundary (matching) or a low-density subset from another unknown shape (non-matching). Density of the comparison shapes was varied across four levels - 4, 8, 16, and 32%. Respondents saw a given target or non-matching comparison shape only once. An example of each density level is provided in the panels of Figure 2.

Respondents were instructed to decide whether the comparison shape had been derived from, and thus appeared similar to, the target shape. They were told to answer "same" or "different" to register this judgment. This was the same experimental protocol used by Greene & Hautus [17], except that they used targets with 100% density and here the targets themselves were shown with low-density (sparse) subsets of boundary dots. Additional details on equipment and task conditions can be found

Eight respondents provided judgments for this task, consisting of 160 matching trials and 160 non-matching trials. Figure 3 shows a plot of the percent of matching judgments that were correct. The differentials produced by the 4% and 32% treatments were each significant at $p < 0.0001$, as indicated by one-way Anovas with repeated measures. A two-way repeated measures Anova across both of these treatments did not find the main effects to be significant. The interaction of target density by comparison density was significant at $p < 0.0001$.

Clearly the two levels of target density had a reciprocal influence on similarity judgments. For 32% target patterns the probability of same judgments declined as the density of comparison shapes was reduced from 32% to 4%. For 4% targets the probability of same judgments declined as the density of comparison shapes was increased from 4% to 32%. This demonstrates that shape similarity is not based on enumerating the number of matching boundary markers, nor are the dots of the target pattern delivering a modicum of shape information

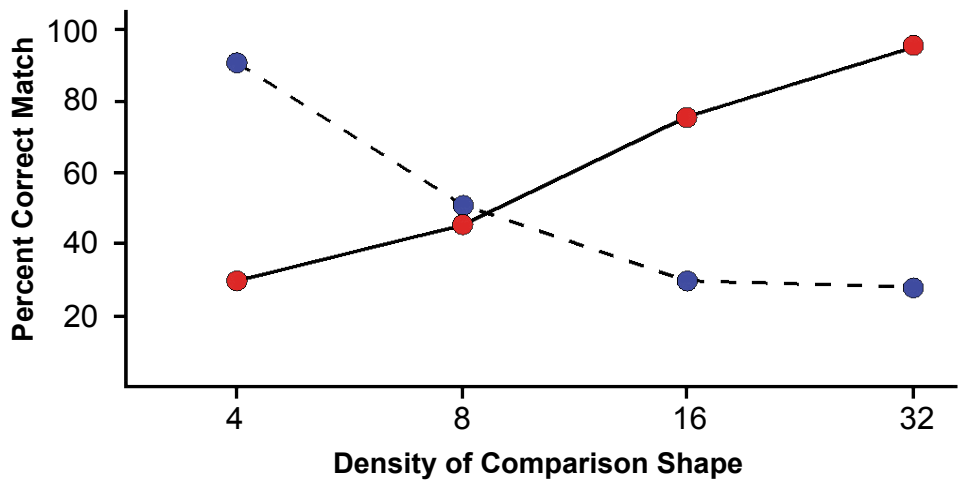

**Figure 3:** Mean hit rates for matching judgments are plotted. Judgments on trials that used 32% targets are shown with a solid line and those from the 4% targets are shown with a dashed line.

It seems likely, in retrospect, that our visual system can derive relationships among the dots from a low-density target, providing what can be described as a shape summary. A low-density sample of dots from the original unknown shape provides a shape perception that differs from what the full boundary will generate. Adding dots will conflict with that summary, decreasing the likelihood that it will be judged as a match. Dot locations are the elemental shape cues, with their relationships providing the summary that we store in working- or long-term memory for purposes of identification. The lesson here is that even a stimulus having very few dots can generate shape information that dominates perceptual judgments. Our simplest shape concept is provided by three dots, which is perceived as a triangle - see [18] for additional discussion of this issue.

Note that we are making no distinction between the concept of a shape and that of a pattern. The latter term is commonly used in describing a small set of discrete spatial locations that may not have specific alignments such as those found in a contour. Nonetheless, it is likely that the basic encoding mechanism for registering shape boundaries is the same whether one is using discrete dots or a continuous contour. Both provide location markers and most of the markers in a continuous boundary deliver redundant information.

It is significant that here, as in the earlier study [17], matching judgments were made within a few seconds after a single exposure of a given unknown shape. This reinforces the argument that connectionist models are not providing a valid concept for how shapes are encoded because they require many hundreds or thousands of trials to achieve shape encoding. See [17] for further discussion of that issue.

that V1 neurons are not providing the most elemental shape cues. For the unknown shape inventory, a 4% density has spans between dots that are generally greater than the length of the receptive fields of V1 neurons. Shapley's lab [19] examined the receptive fields of orientation-selective neurons in V1 of Macaque, which is thought to be functionally similar to primary visual cortex in humans. For 30 out of 31 neurons the length of the excitatory zone was less than 2.5 arc°. The 4% displays provided mean separation of 3.8 arc° for both horizontal and vertical spans, and a separation of 5.4 arc° for dot-pairs lying at 45° orientations. Seventy-eight percent of the horizontal and vertical spans were over the 2.5 arc° receptive-field length of V1 neurons.

To further examine this issue, the test trials that displayed both target and matching shapes at 4% density were sorted to identify shapes wherein all dot separations were greater than 2.5 arc°. Those shapes were identified correctly on 91% of the trials - the same hit rate as for shapes that included shorter separations. For the shapes having only large dot separations, the V1 neurons would register only a single dot, which provides no information about orientation, length, or curvature. We conclude that the elemental cues for shape recognition are the marked boundary locations, from which the system derives relative distance and angle information that defines the shape or pattern. It is misleading to view orientation, length, and curvature as elemental.